\documentclass[submitting]{nst}

\usepackage{subfigure,dcolumn}
\usepackage[T2A,T1]{fontenc}
\usepackage[russian,english]{babel}

\usepackage{listings}
\usepackage{makecell}
\usepackage{subfigure}

\newcommand{\msr}{$\mathrm{\mu}$SR\ }

\begin{document}

\title{Test of LGAD as Potential Next-Generation \msr Spectrometer Detectors}\thanks{Supported by the National Key Research and Development Program of China (No.2023YFE0105700); by Basic Frontier Scientific Research Program of the Chinese Academy of Sciences (No.ZDBS-LY-SLH009); by National Natural Science Foundation of China (No.12175252).}

\author{Yu-Hang Guo}
\affiliation{Spallation Neutron Source  Sciences Center, Dongguan 523803, China}
\affiliation{Institute of High Energy Physics, Chinese Academy of Sciences, Beijing 100049, China}
\author{Qiang Li}
\email[Corresponding author, ]{Qiang Li. address: Zhongziyuan Road 1st, Dongguan 523803, China. phone: 0769-88931196. email: qiangli@ihep.ac.cn}
\affiliation{Spallation Neutron Source  Sciences Center, Dongguan 523803, China}
\affiliation{Institute of High Energy Physics, Chinese Academy of Sciences, Beijing 100049, China}
\author{Yu Bao}
\email[Corresponding author, ]{Yu Bao. address: Zhongziyuan Road 1st, Dongguan 523803, China. phone: 0769-88931976. email: yubao@ihep.ac.cn}
\affiliation{Spallation Neutron Source  Sciences Center, Dongguan 523803, China}
\affiliation{Institute of High Energy Physics, Chinese Academy of Sciences, Beijing 100049, China}
\author{Zi-Wen Pan}
\affiliation{State Key Laboratory of Particle Detection and Electronics, University of Science and Technology of China, Hefei 230026, China}
\author{You Lv}
\affiliation{Spallation Neutron Source  Sciences Center, Dongguan 523803, China}
\affiliation{Institute of High Energy Physics, Chinese Academy of Sciences, Beijing 100049, China}
\author{Rhea Stewart}
\affiliation{ISIS Pulsed Neutron and Muon Source, STFC Rutherford Appleton Laboratory, Didcot OX11 0QX, UK}
\author{Adrian D. Hillier}
\affiliation{ISIS Pulsed Neutron and Muon Source, STFC Rutherford Appleton Laboratory, Didcot OX11 0QX, UK}
\author{Stephen P. Cottrell}
\affiliation{ISIS Pulsed Neutron and Muon Source, STFC Rutherford Appleton Laboratory, Didcot OX11 0QX, UK}
\author{Peter J. Baker}
\affiliation{ISIS Pulsed Neutron and Muon Source, STFC Rutherford Appleton Laboratory, Didcot OX11 0QX, UK}
\author{James S. Lord}
\affiliation{ISIS Pulsed Neutron and Muon Source, STFC Rutherford Appleton Laboratory, Didcot OX11 0QX, UK}
\author{Lei Liu}
\affiliation{Spallation Neutron Source  Sciences Center, Dongguan 523803, China}
\affiliation{Institute of High Energy Physics, Chinese Academy of Sciences, Beijing 100049, China}
\author{Zhi-Jun Liang}
\affiliation{Institute of High Energy Physics, Chinese Academy of Sciences, Beijing 100049, China}
\author{Yun-Yun Fan}
\affiliation{Institute of High Energy Physics, Chinese Academy of Sciences, Beijing 100049, China}
\author{Meng-Zhao Li}
\affiliation{Spallation Neutron Source  Sciences Center, Dongguan 523803, China}
\affiliation{Institute of High Energy Physics, Chinese Academy of Sciences, Beijing 100049, China}
\author{Mei Zhao}
\affiliation{Institute of High Energy Physics, Chinese Academy of Sciences, Beijing 100049, China}
\author{Gao-Bo Xu}
\affiliation{Institute of Microelectronics, Chinese Academy of Sciences, Beijing 100029, China}
\author{Mei-Chan Feng}
\affiliation{Spallation Neutron Source  Sciences Center, Dongguan 523803, China}
\affiliation{Institute of High Energy Physics, Chinese Academy of Sciences, Beijing 100049, China}

\begin{abstract}
 Muon Spin Rotation/Relaxation/Resonance (\msr) is a versatile and powerful non-destructive technology for investigating the magnetic properties of materials at the microscopic level. 
The \msr technique typically utilizes fully spin polarized beams of positive muons generated at particle accelerator facilities and measures the evolution of the muon spin polarization inside a sample to extract information about the local magnetic environment in materials.
With the development of accelerator technologies, intensities of muon beams are being continuously improved, which will cause a pile-up problem to the \msr spectrometer.
This problem is becoming especially challenging at intense pulsed muon sources since the instantaneous data rates are expected to be much higher.
The first muon source in China, named MELODY, is currently under construction and will be a pulsed source of muons operated at a repetition frequency of only 1 Hz due to limitations of the accelerator system at CSNS. 
Consequently, there is a strong motivation to operate MELODY at significantly higher muon intensities. 
This necessitates an upgrade of the detector system inside the spectrometer, which should be smaller and faster to accommodate the increased intensity per pulse of muons. 
The Low Gain Avalanche Diode (LGAD), characterized by a typical pulse width of 2 ns and a segmentation size in the centimeters range, has the potential to significantly improve the counting rates of \msr spectrometers that utilize a high intensity pulsed muon source.
Thus, it is expected that the LGAD detector is a promising candidate to enhance the performance of \msr spectrometers at the new MELODY muon source. 
To validate this, tests on the LGAD were conducted at the ISIS pulsed muon source at the Rutherford Appleton Laboratory, UK. 
This paper will describe the setup of the candidate LGAD devices and the subsequent analysis of the experiment data.
\end{abstract}


\maketitle

\section{Introduction}

\label{sec:intro}
Non-Destructive Technologies (NDT) are a class of powerful experimental techniques used to investigate the properties of materials. 
Nuclear Magnetic Resonance (NMR) and Electron Spin Resonance (ESR) techniques, which are representative NDTs, usually require both high magnetic fields and low temperatures to produce a thermal equilibrium spin polarization environment. 
NMR also requires specific target nuclei. 
Thus, their application is often restricted. By contrast, the Muon Spin Rotation/Relaxation/Resonance (\msr) is relatively versatile \cite{muSRBook-PSI,muSRBook-JPARC,muSRBook-ISIS,muSRBook-ISIS-np}.

The muon is a good and complementary probe of magnetism local to where it stops in materials \cite{muSRBook-PSI}.
In the application of the \msr technique, fully spin polarized positively charged muons ($\mu^+$), which are produced by an accelerator muon source, are implanted into a sample \cite{muSR1,muSR2,muSR3}. Upon implantation, the muons rapidly thermalise before their spins begin to precess about the local magnetic field at the muon stopping site, which is characteristic of the magnetic properties of the material being studied \cite{muSR3}. The muon is an unstable particle and with a lifetime of $\sim2.2$ $\mu$s. 
The $\mu^+$ will decay into a positron and two neutrinos.
Since the decay positrons are preferentially emitted along the momentary muon spin direction at the point of decay, the precession of the muon spin polarization can be deduced by comparing the number of emitted positrons detected at different angles around the sample. From this signal it is then possible to characterize in detail the local magnetic fields that the muon spins experienced and this can reveal information about magnetic, superconducting and dynamic properties of materials on the microscopic level. Based on the observation of the time evolution of the implanted muon spin polarization, the \msr technique provides a versatile method to perform measurements without the need to apply an external magnetic field, which makes it an essential tool in many frontier fields including material science, condensed matter physics, and chemistry \cite{muSR1,muSR3,muSRBook-mag,muSRBook-che}.

\msr experiments are strongly dependent on the muon source used, which can usually be classified into one of two types, continuous wave (CW) muon sources and pulsed muon sources, which offer complimentary strengths for application to \msr. 
On a CW muon source, such as the CMMS facility at TRIUMF \cite{TRIUMF}, the S$\mu$S facility at PSI \cite{PSI} and the MuSIC facility at RCNP \cite{RCNP}, \msr experiments usually have good time resolution to the level of nano-seconds or less, but the statistics are count rate limited because there is only a single muon in the sample at a time \cite{muSRBook-PSI}.
On a pulsed muon source such as the ISIS facility at RAL \cite{RAL} and the MUSE facility at J-PARC \cite{JPARC}, \msr instrument count rates are much higher and measurement backgrounds are lower, which means good statistics can be gathered even at longer times. 
However, the time resolution, and therefore the maximum strength of magnetic field the muons can measure, is set by the width of the muon pulse, which is typically of the order of 100 ns. This means for a given experiment it can often be useful to measure using both continuous and pulsed muon beams.

The first surface muon source in China (called MELODY, short for the Muon Station for Science Technology and Industry) \cite{MELODY} is currently under construction at the campus of China Spallation Neutron Source (CSNS) \cite{CSNS,CSNS-NAT}. 
The MELODY is designed to provide a pulsed muon beam with an intensity in the range of $10^5$ muons per pulse, and retains the ability to increase the intensity to the level of $10^6$ muons per pulse or more with future upgrades. 
However, limited by the operation plan of the CSNS accelerator, the frequency of MELODY is only 1 Hz. 
This will greatly limit the statistics of \msr experiments \cite{MELODY-muSR-2024} and gives great motivation to running MELODY at a higher intensity.
But a higher muon intensity per pulse will also cause a pile-up problem in the detectors.
The decay positrons will be incident into the detector channels closely spaced in time, resulting in challenges in accurately measuring the time of individual positron events.
Therefore, to utilize a higher intensity muon source, the detectors used in \msr spectrometers need to be faster and more highly segmented. 
A smaller detector pixel size decreases the possibility of multiple decay positrons reaching a single detector at overlapping times, while a faster detector with a narrower signal width increases the likelihood of distinguishing the arrival time of signals accurately. 
The dominant detector technology applied in recently developed \msr spectrometers is the "scintillator with SiPM" system, which usually has a signal pulse width on the order of 100 ns and a detector pixel size of several squared centimeters.
The research group at MELODY is designing a \msr spectrometer following the "scintillator with SiPM" method \cite{MELODY-muSR-2024,MELODY-muSR-2023,MELODY-muSR-2022,MELODY-muSR-2021,MELODY-muSR-2019,MELODY-muSR-iop}. 
Many efforts have been contributed to study the performance of this spectrometer running on MELODY.
The result shows that a \msr spectrometer based on a scintillator with SiPM detector system cannot utilize the full muon flux per pulse available at MELODY and would result in a waste of muon beam resource.

A novel type of detector, the Low Gain Avalanche Diode (LGAD) \cite{LGAD-0}, is a promising candidate system to solve this problem.
The LGAD is an advanced detector technology that has been developed in recent years to measure the arrival time of particles at the end cap of ATLAS and CMS under the heavy irradiation on LHCb. 
During the progress of development, the research group from China has gained many achievements: 
The group from IHEP has presented the design of the LGAD with shallow carbon and deep $N^{++}$ layer \cite{LGAD-irr-0,LGAD-irr}, which has greatly improved the radiation hardness to $2.5\times10^{15} n_{eq}/cm^2$.
They also contributed many efforts to improve the performance of LGAD \cite{LGAD-1,LGAD-2,LGAD-3,LGAD-4,LGAD-5}.
Meanwhile, group from USTC of China has conducted studies on the dynamic properties of LGAD \cite{LGAD-USTC}. 

The notable features of the LGAD are listed in table \ref{tab:lgad}.
Given the small width of signal pulse without detector dead time, the response of the LGAD to charged particles is very fast.
The fast response, as well as the small segmentation size, are beneficial to improving the counting rate of \msr spectrometers and address some of the limitations with scintillator-SiPM based technologies.
In addition, the LGAD's thin active volume will reduce the effect of multi-counting among neighboring channels in a detector matrix and the good irradiation hardness makes it work stably, especially under the high radiation environment at a high intensity muon source.
Based on these features of detection, the LGAD is expected to be a good candidate for replacing the scintillator technology that is widely applied in existing \msr spectrometers and will improve the count rate capability by utilizing an increased muon source flux. 

\begin{table}[h]
\centering
\caption{Selected performance parameters of LGAD detector \cite{LGAD-irr}.}
\begin{tabular}{c|c}
\Xhline{1.2pt}
Performance              & Parameters           \\
\Xhline{1.2pt}
segmentation size & $1.3\times 1.3~\mathrm{mm^2}$ \\
\hline
gain                     & 10$\sim$30           \\
\hline
active thickness          & 50 \textmu m           \\
\hline
timing resolution      & 30 ps                \\
\hline
full pulse width      & 2-3 ns                \\
\hline
irradiation hardness      & $>10^{15} n_{eq}/cm^2$                \\
\Xhline{1.2pt}
\end{tabular}
\label{tab:lgad}
\end{table}

To demonstrate this, a test experiment has been conducted at the ISIS pulsed muon source at RAL, which will be discussed in this paper. The instrumentation used for this test experiment is described in chapter \ref{sec:setup}, including the characteristics of the muon source, the CHRONUS muon instrument and the detector setup. 
In chapter \ref{sec:ana}, a summary of the data, the progress of the data analysis and a discussion of the results is presented.
Finally, in the last chapter, the experiment and its conclusions will be summarized and the application of LGAD in \msr technology will be prospected.

\section{Instrumentation Setup}
\label{sec:setup}
The test experiment introduced above was conducted using the ISIS pulsed muon source at RAL in the UK. The LGAD detector was produced in China, and a test board carrying the LGAD detector and the readout electronics was assembled in China and transported to the experiment at RAL. 
Details about the instrumentation used for this experiment will be discussed in this chapter.

\subsection{The Muon Source}
\label{sec:source}

\begin{figure}[h]
\centering
\includegraphics[width=.4\textwidth]{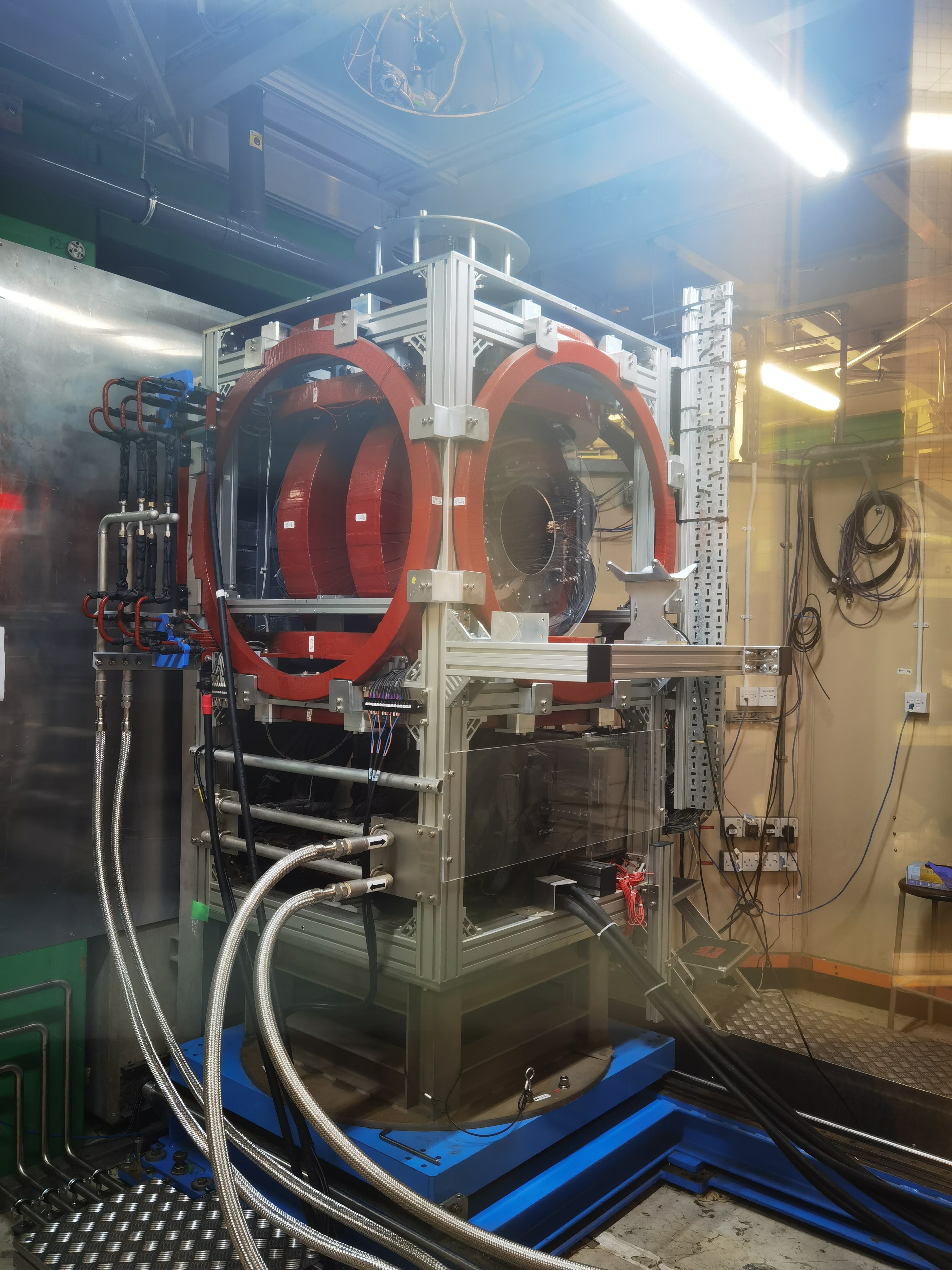}
\caption{Photo of CHRONUS. The coils are fixed by a Bosch framework and are able to provide uniform magnetic fields of up to 3950 G. \label{fig:chronus}}
\end{figure}

The experiment utilized the surface muon beam delivered to the CHRONUS instrument port, which is one of the suite of muon spectrometers at ISIS \cite{CHRONUS,RIKEN-RAL}.
Fig.\ref{fig:chronus} shows a view of the CHRONUS instrument. 
It has 606 detectors and can be used for zero field (ZF), transverse field (TF) and longitudinal field (LF) \msr experiments. 
The magnetic fields are produced by the red magnet coils that are fixed onto Bosch framework. Fields of up to 3950 G can be applied. During our test, TFs of 0 Gauss, 50 Gauss and 100 Gauss were applied during the various tests conducted.

The beam snout through which the muons are delivered to the experiment is obscured by the CHRONUS instrument in Fig.\ref{fig:chronus}. The muon beam is produced upstream of the beam port, where 800MeV protons are incident on a carbon target and the resultant muons extracted and transported downstream to the experiment. 
Depending on the features of the proton beam, the extracted muon beam is provided at a repetition frequency of 40 pps, with the FWHM of each pulse around 55 ns \cite{CHRONUS}.
Although the muons are produced in double bunch mode, they can be delivered to CHRONUS as single or double pulses \cite{RIKEN-RAL}. For this experiment, CHRONUS was configured for single bunch mode operation, which was most suitable for testing the LGAD system. This experiment was performed in parallel with a second set of experiments on a scintillator based detector system, which is being developed for the new \msr spectrometer on MELODY \cite{MELODY-muSR-2024}. The intensity of the incident muon beam was in the range of $10^4$ to $10^5$ muons per second.

A T0 signal is available from the proton accelerator 4 $\mu$s in advance of the beam pulses arriving at the muon production target.
This signal is used as a time reference to start monitoring the time evolution of muon spin polarization, which holds important information about the magnetic properties of the sample to be measured in \msr experiments.

\subsection{The Detector Setup and the devices}

\begin{figure}[]
    \centering
    \subfigure[]{    
		\label{fig_device}     
	    \includegraphics[width=0.3\textwidth]{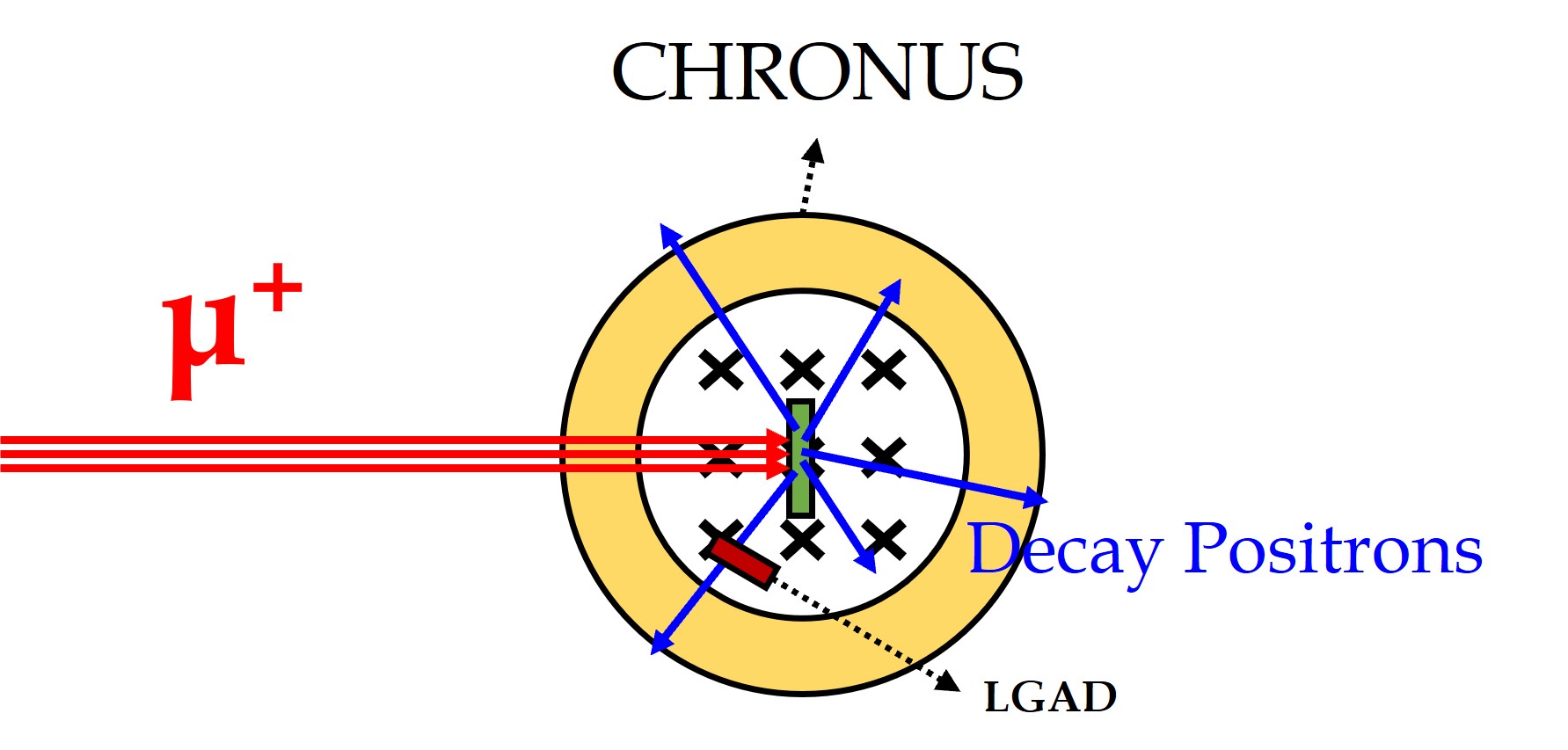}}
    \qquad
    \subfigure[]{    
		\label{fig_field}     
		\includegraphics[width=0.3\textwidth]{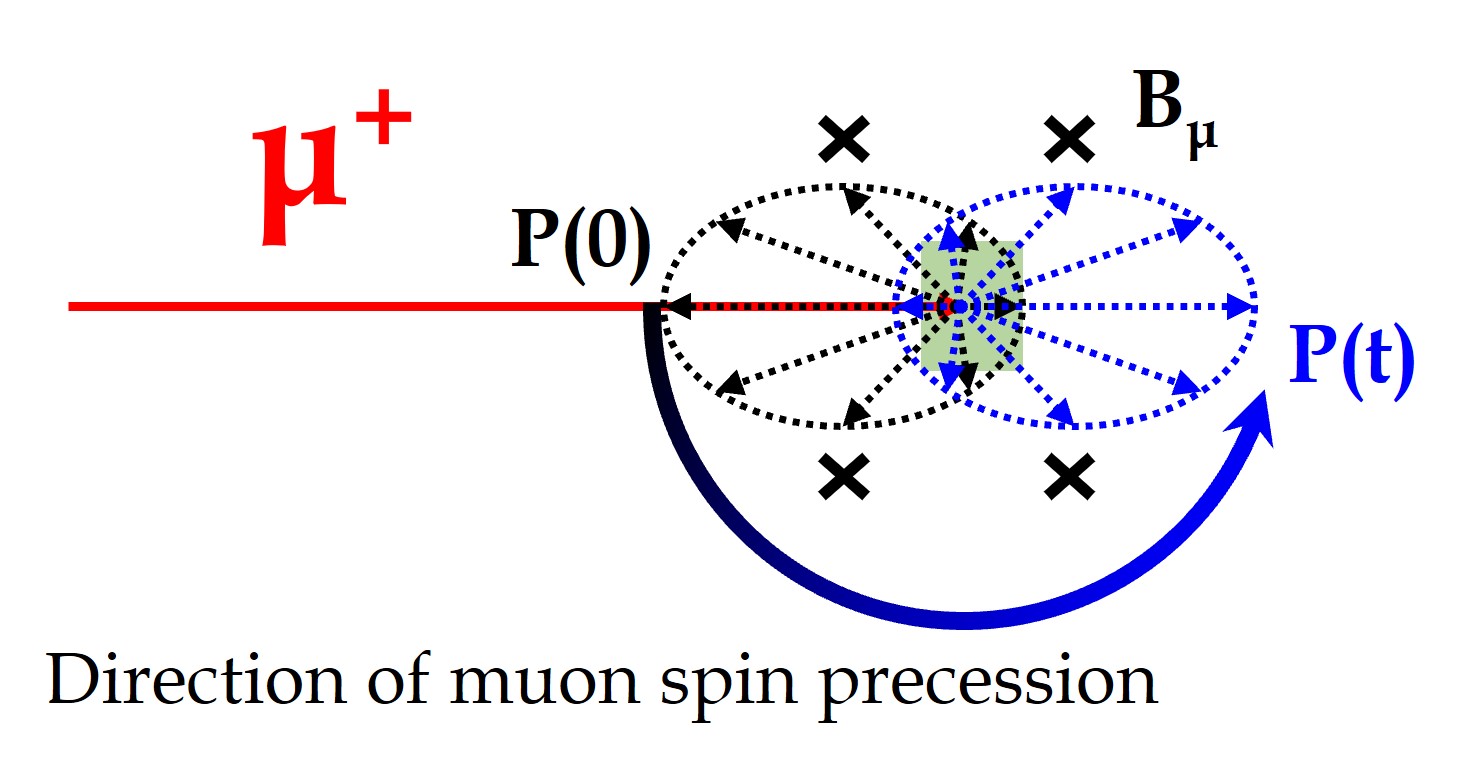}}
      \qquad
    \subfigure[]{    
		\label{fig_precession}     
		\includegraphics[width=0.3\textwidth]{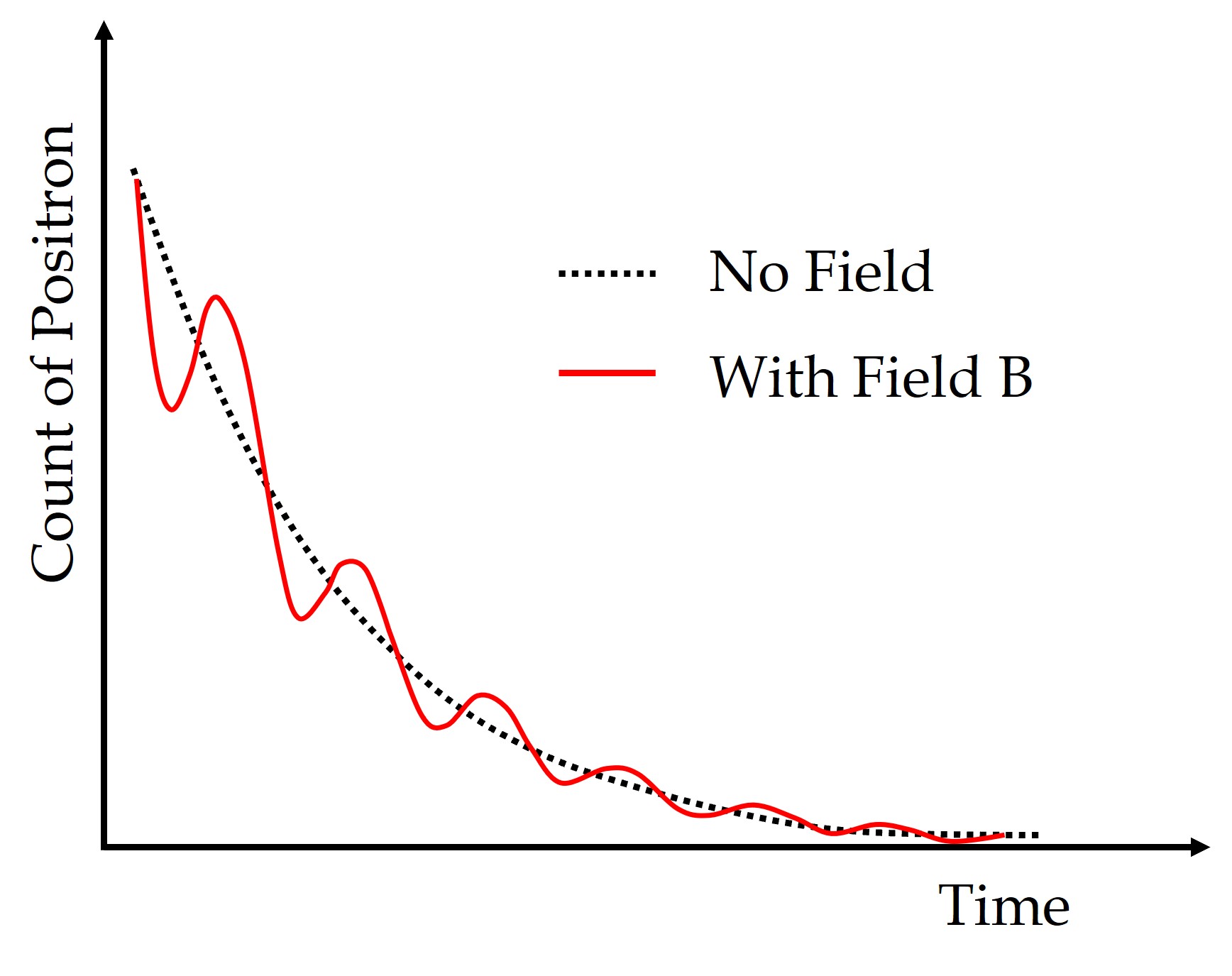}}
	\caption{(a)Sketch of the horizontal profile of the device setup (not to scale). The green block indicates the sample in the muon beam. The yellow ring indicates the magnetic field coils of CHRONUS, which generate a magnetic field along the perpendicular direction with respect to the plane of the paper. The LGAD is fixed to the side of the sample. The positrons are emitted when muons that stop in the sample decay and the fraction of these that intersect the LGAD device will be detected. (b) Sketch of the field and the expected evolution of the muon spin polarization. The precession of the muon will influence the count of positrons detected by the LGAD. (c) Arbitrary demonstration of the \msr signal influenced by the precession of the muon. Oscillations coupled to the exponential function are due to the Larmor precession of muons about a net field. Details see text.}
	\label{fig:setup}
\end{figure}

The layout of the detector setup can be found in the sketch shown in Fig.\ref{fig_device}.
The yellow ring indicates the magnet coils of the CHRONUS spectrometer, which generate a TF along the perpendicular direction with respect to the plane of the paper.
The sample is represented by the green block and placed in the path of the muon beam (as indicated by the red lines).
The material of the sample is silver which is paramagnetic and has a very weak nuclear moment. Therefore, the field inside the silver sample is approximately equal to the field generated by CHRONUS.
The muon beam is incident from the left and is implanted in the silver sample. Since the kinetic energy of the surface muons is low, all the muons will be stopped in the sample and then decay into positrons and neutrinos. 

The muons provided by the source are to good approximation 100\% spin polarized, with the initial muon spin direction aligned opposite to the muon momentum.
Because the decay positrons are preferentially emitted along the direction of muon spin at the moment of decay, when the muons stop in the material, the initial direction of decay positrons emitted will be distributed as shown by the black dashed line in Fig.\ref{fig_field}. 
In the presence of a magnetic field in the sample, the spin of the stopped muons will rotate under Larmor precession. 
After time t, the average direction of the muon spins rotates to the right, and the direction of decay positrons emitted will be distributed as shown by the blue dashed line.
When a detector is fixed beside the sample, the positron counts detected as a function of time will be influenced by the muon spin rotation.
Where the muons precess in phase in a net magnetic field, an oscillation will be found coupled to an exponential signal that arises due to the muon lifetime as shown in Fig.\ref{fig_precession}.
The frequency of the oscillation is dependant on the magnetic field strength, whereas the amplitude is dependent on the magnetic volume fraction, detector location and any degrader\cite{muSRBook-ISIS}.

Based on this principle, the LGAD is fixed in a plane to the side of the sample.
Its zenith angle to the muon beam is deviated from 90 degrees to avoid the LGAD being hit by positrons present in the incident beam which miss the sample or are scattered by it
A signal will be generated and read by the readout system when a positron hits the sensitive detection area of LGAD.

\begin{figure}[]
    \centering
    \subfigure[]{    
		\label{fig_s}     
	    \includegraphics[width=0.4\textwidth]{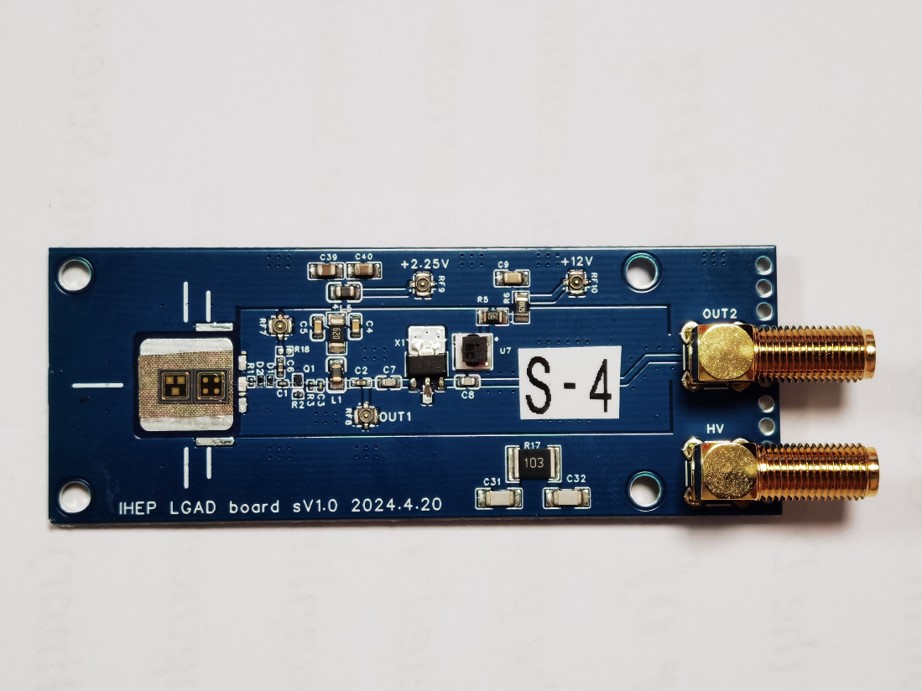}}
    \qquad
    \subfigure[]{    
		\label{fig_t}     
		\includegraphics[width=0.4\textwidth]{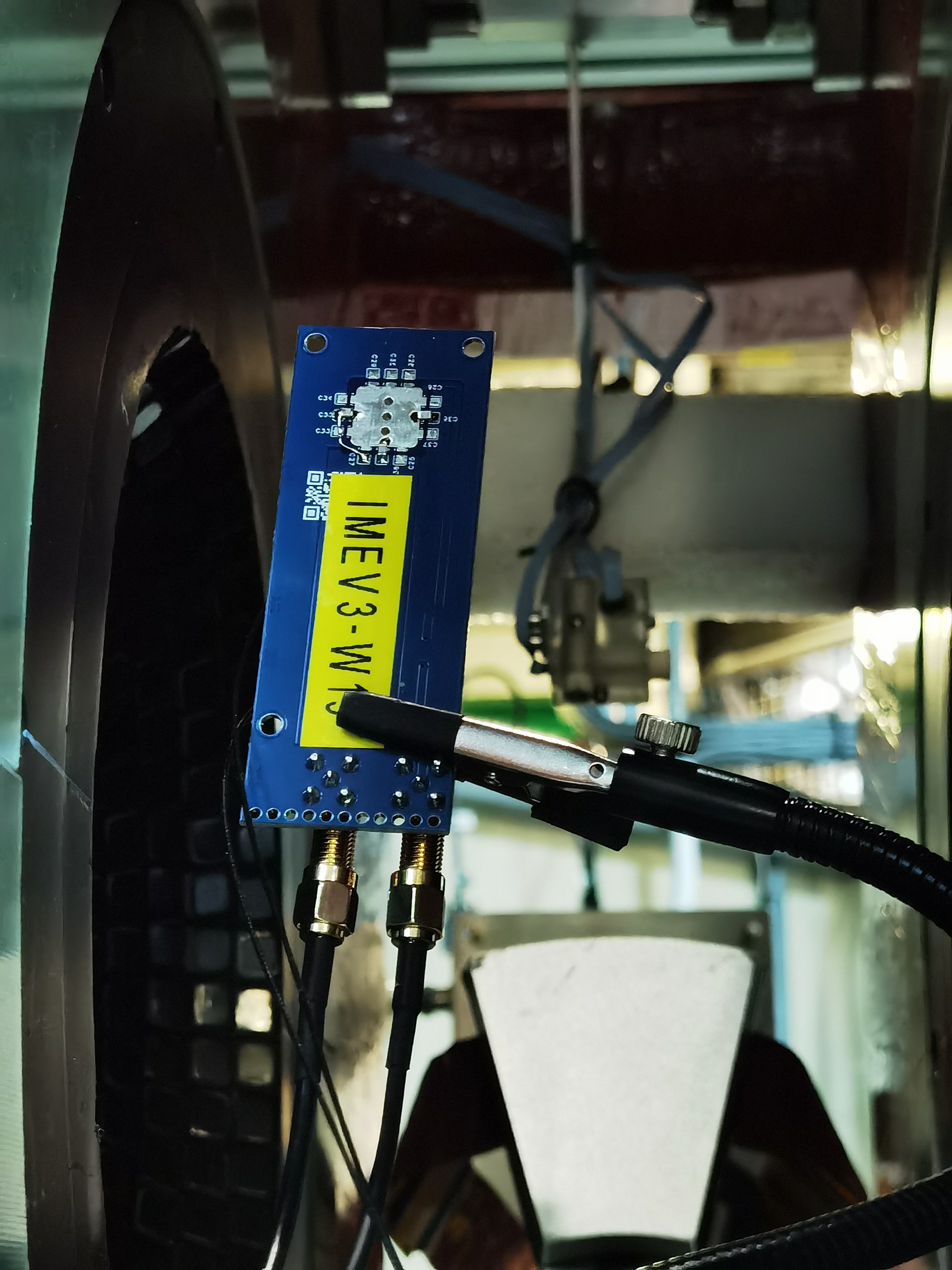}}
	\caption{Photos of the LGAD detector under test. An LGAD of size $2.6\times2.6 mm^2$ is bonded on the detector PCB (a). The detector PCB is fixed by a clamp (b). The detectors used in this test are produced by the Institute of High Energy Physics, CAS. The yellow label in (b) does not represent for any real information.}
	\label{fig:det}
\end{figure}

The LGAD detector is bonded on a PCB, as shown in Fig.\ref{fig:det}. 
The sensitive area is a $2.6\times2.6\ \mathrm{mm^2}$ square on the left part of the device as shown in Fig.\ref{fig:det}(a). 
The remaining area on the PCB is occupied by the amplifier circuits. 
Signals from the LGAD after the amplifier are read out from the SMA connector at the right edge of the PCB. 
The detector PCB is fixed by a clamp as shown in Fig.\ref{fig:det}(b), with its sensitive area facing towards the sample in the beam.
The distance from the sample to the detector is approximate to 15 cm.

\subsection{The Readout System}

\begin{figure}[h]
    \centering
    \subfigure[]{    
		\label{fig_frosa}     
	    \includegraphics[width=0.4\textwidth]{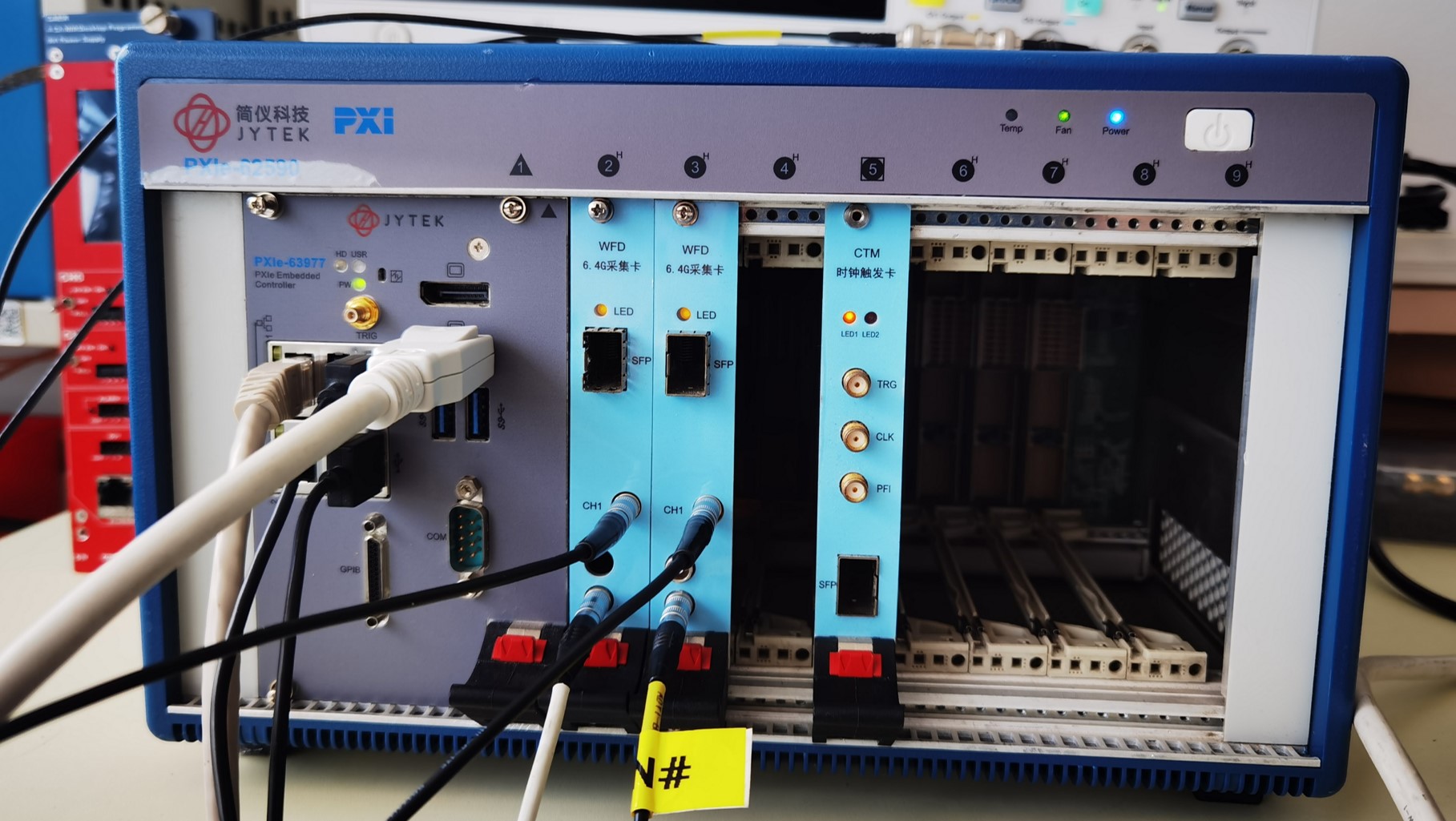}}
    \qquad
    \subfigure[]{    
		\label{fig_frosb}     
		\includegraphics[width=0.4\textwidth]{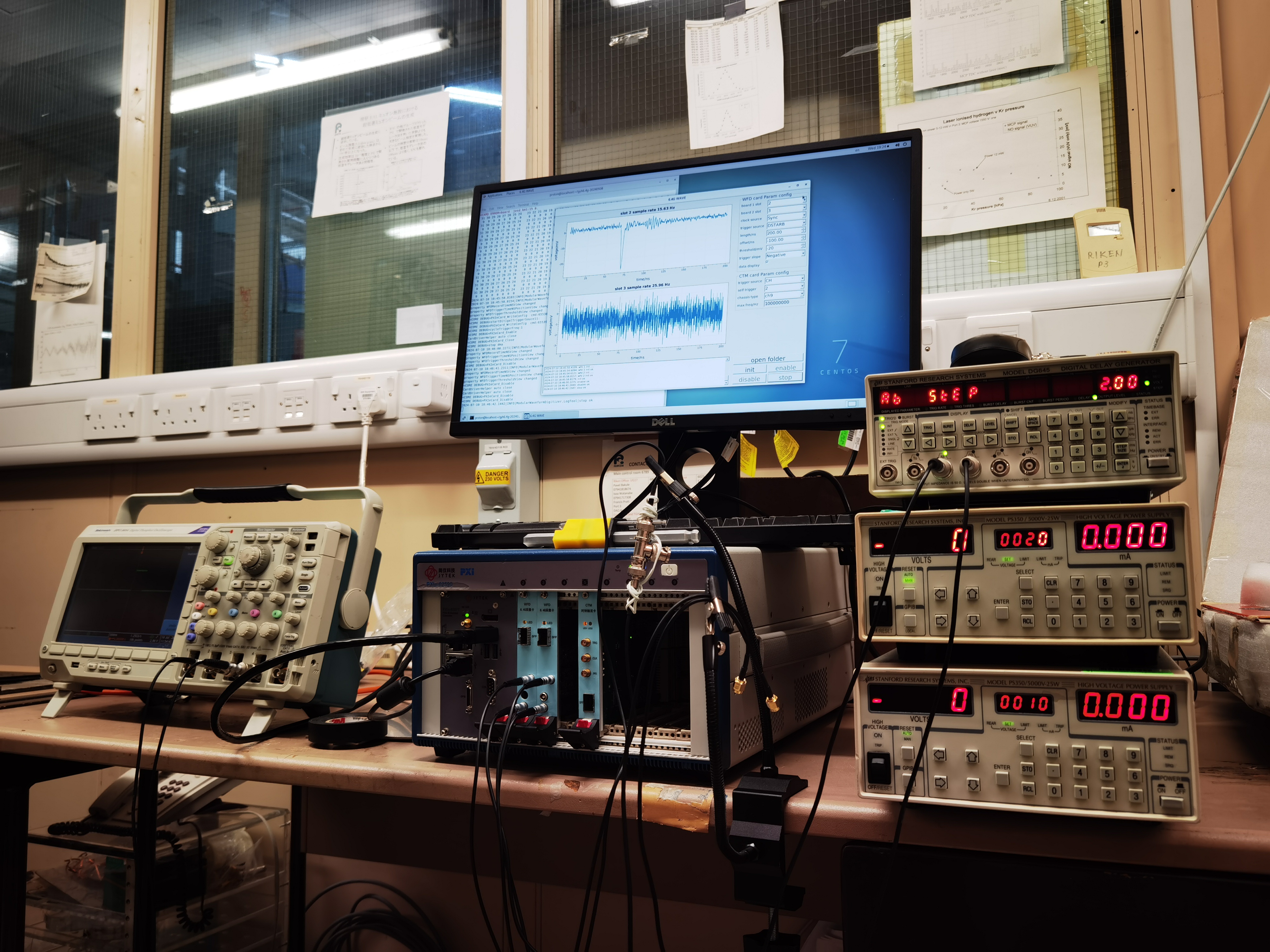}}
	\caption{(a) Photo of the FROS. Two channels with 6.4 Gsps sampling rate are loaded. (b) Photo of FROS under test in lab. Thanks to the support of power supplies provided by ISIS muon group.}
	\label{fig:fros}
\end{figure}

The signal of the LGAD is very fast. 
Usually, the width of the rising edge is around 700 ps and the full width of a signal is around 2 ns. 
Because of the very fast signal, a readout system with high sampling rate is needed.
With this in mind, the Fast Read-Out System (FROS) \cite{FROS} was taken to ISIS for the experiment to be used to read and store the signals generated by the LGAD detector.
Photos of the FROS can be found in Fig.\ref{fig:fros}.
In total, two DAQ channels were loaded, making the system able to read signals from two detector synchronously. 
(However, only one LGAD was used and connected to one of the DAQ channels in the FROS.)
Beside the two DAQ channels, a controller in charge of DAQ control and data storage and a TCM board in charge of multi-channel clock alignment, were loaded.
The T0 signal introduced in section \ref{sec:source} was connected to the FROS. 
The timing clock on the DAQ channels was initiated as soon as the FROS received a T0 pulse. The voltage trigger threshold and the length of data-acquisition were adjustable. The FROS records the waveform it receives immediately when the voltage threshold is triggered, as well as the time of trigger ($T_s$) since the most recent T0 pulse.
From $T_s$, the measured sample signal for a \msr experiment can be extracted after the necessary analysis.

\section{Analysis and Result}
\label{sec:ana}

The progress of the data analysis will be introduced in this chapter.
In the first section, a summary of the data will be provided and the waveform recorded will be displayed and discussed. 
In the second section, the analysis of the \msr signals, which hold the information about the evolution of the muon spins will be described. 
Lastly, the performance of the LGAD in this test and the future prospects for its application to \msr spectrometers will be discussed.

\subsection{Waveform of Positron Signals}

The statistics for the test of the LGAD are summarized in table \ref{tab:time}.
The active time of data acquisition was around 31.4 hours in total.
During the test, the event rate was kept to a stable and low level, which was approximately 2.3 $\mathrm{event/s}$ on average.

\begin{table}[t]
\centering
\caption{Summary of the statistics of the test}.
\begin{tabular}{c|c|c|c}
\Xhline{1.2pt}
TF (G)    & DAQ time (h)    & Count of events     & rate of events (per second)           \\
\Xhline{1.2pt}
0 & 4.1 & 33273 & 2.26  \\
\hline
50 & 15.8 & 131018 & 2.30  \\
\hline
100 & 11.5 & 96141 & 2.32  \\
\Xhline{1.2pt}
\end{tabular}
\label{tab:time}
\end{table}

\begin{figure}[h]
    \centering
    \subfigure[]{    
		\label{fig_wavea}     
	    \includegraphics[width=0.5\textwidth]{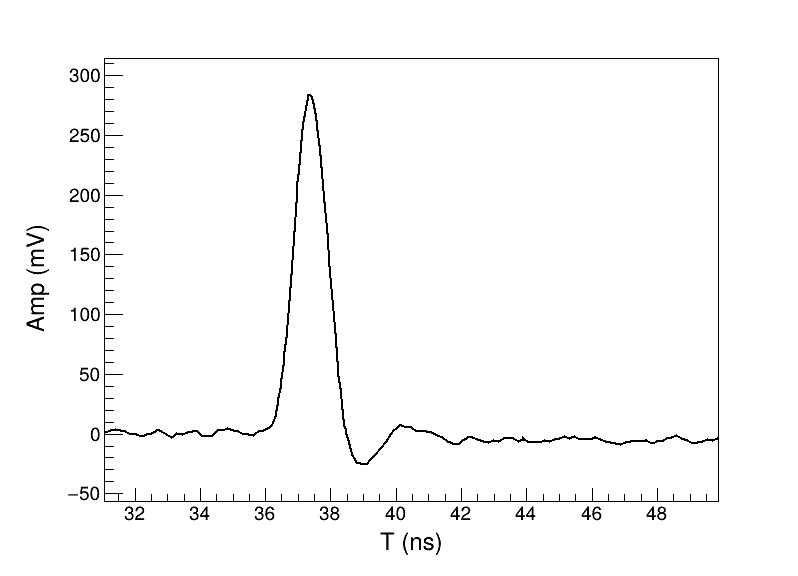}}
    \qquad
    \subfigure[]{    
		\label{fig_waveb}     
		\includegraphics[width=0.5\textwidth]{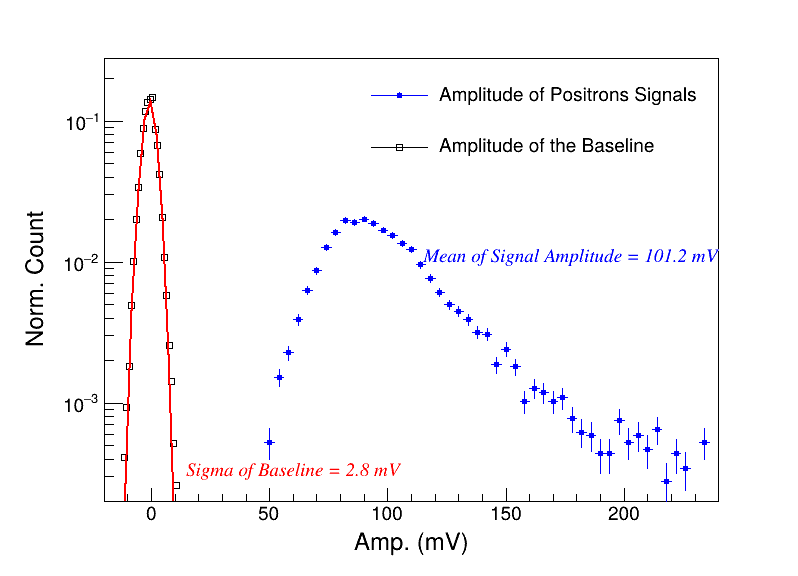}}
	\caption{(a) Demonstration of one of the waveforms recorded by the FROS. The full width of the waveform is around 2 ns. The white noise on the waveform baseline is relatively weak. (b) Comparison between the amplitude of positron signals (blue markers) and white noise (black opened markers). A threshold of 50 mV was applied to FROS during the test. With Gaussian fitting, the sigma of the amplitude of the white noise is around 2.8 mV, whereas the mean of the signal amplitude is more than 101 mV. The signal to background ratio is more than 30, which implies that the signals of muon decay positrons on LGAD is significant enough. details see text.}
	\label{fig:wave}
\end{figure}

One of the concerns of detecting muon decay positrons with LGAD is the significance of the signal. 
The average energy of the positrons is 29 MeV, which is very fast and in the state of minimum ionization particle (MIP).
Particles in the MIP state usually deposit low energy and make the signal not significant enough to be detected.
To investigate this concern, the waveform of the signals are summarized and demonstrated in Fig.\ref{fig:wave}.
In Fig.\ref{fig_wavea}, it can be seen that the full width of the waveform is around 2 ns, which is the reason for using FROS. 
The signals after the amplifiers have an average amplitude of 101.2 mV, which is illustrated by the blue markers in Fig.\ref{fig_waveb}. 
To evaluate the oscillations of the white noise in the waveform, the amplitudes of the baseline in every waveform are summarized and illustrated by the histogram with the black opened markers in Fig.\ref{fig_waveb}.
This can then be fitted by a Gaussian function, from which sigma of the white noise amplitude is found to be around 2.8 whereas the mean is approximate to 0.
Considering that the signal to background ratio is more than 30, this suggests that the signals of the muon decay positrons on LGAD are significant enough.
This has indicated that the LGAD has the ability to detect the decay positrons of muons.

The time of a positron can be further obtained according to the time over threshold, and is approximately equal to the time of the muon decay since the decay positron is very fast.

\subsection{Result of the Analysis}

\begin{figure}[]
\centering
\includegraphics[width=.5\textwidth]{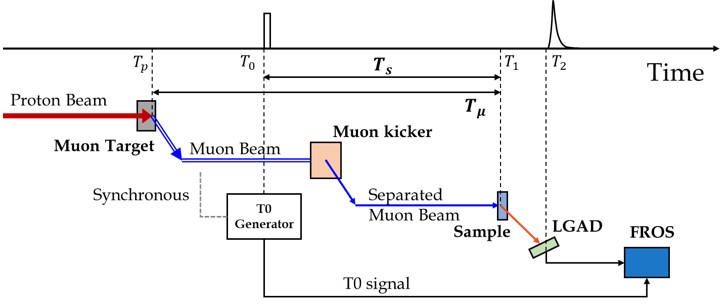}
\caption{Demonstration of the method to achieve the muon decay time. Details see text. \label{fig:timing}}
\end{figure}

The target of a \msr experiment is to acquire the magnetic information about a sample from the evolution of muon polarization, which is characterized by the time of positrons detected by LGAD ($f(t)$).
Thus, $f(t)$ will be measured most directly in a \msr experiment.
The way to achieve this is illustrated in Fig.\ref{fig:timing}.
$T_p$ represents the time when the muon is produced.
$T_1$ represents the time when the decay positron is detected by the LGAD.
$T_0$ represents the time when FROS receives the T0 pulse, which depends on the path length between the muon target and muon kicker, and the length of the T0 signal cable.
During the test, $T_0-T_p$ was always kept constant.

The most ideal situation is to measure $T_1-T_p$.
However, $T_p$ cannot be measured exactly.
Considering that $T_0-T_p$ is fixed and an absolute value of $T_1-T_p$ is not necessary, thus $T_1-T_0$ can be measured to get $f(t)$ in the experiment.

\begin{figure}[t]
    \centering
    \subfigure[]{    
		\label{fig_osca}     
	    \includegraphics[width=0.4\textwidth]{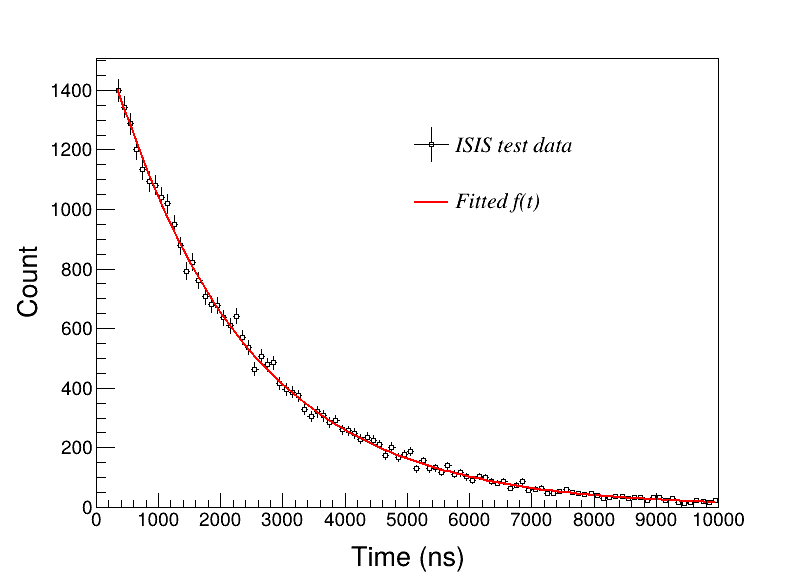}}
    \qquad
    \subfigure[]{    
		\label{fig_oscb}     
		\includegraphics[width=0.4\textwidth]{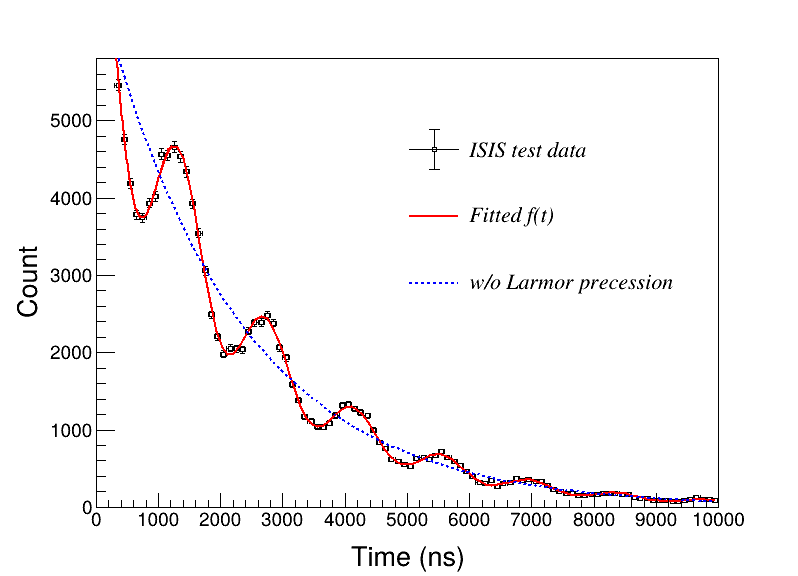}}
      \qquad
    \subfigure[]{    
		\label{fig_oscc}     
		\includegraphics[width=0.4\textwidth]{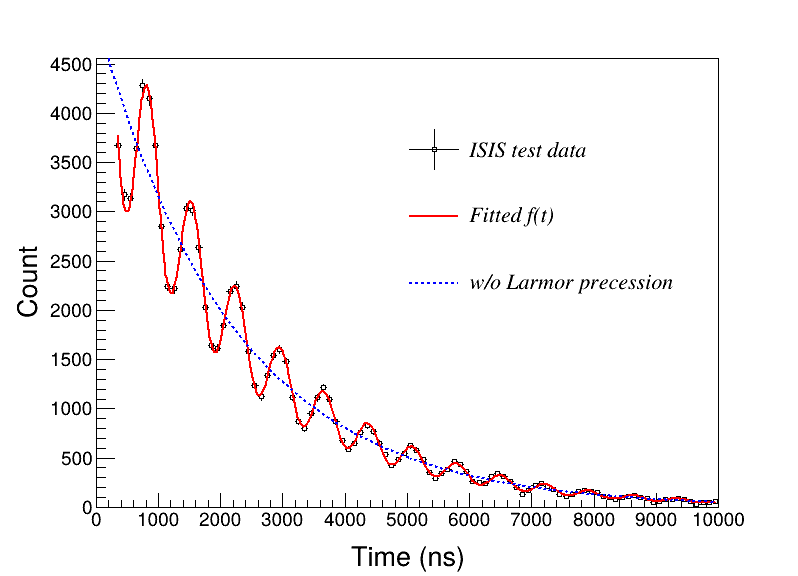}}
	\caption{Result of the analysis. Distribution of the muon decay time. Samples are placed under different magnetic fields of 0 G in plot (a), 50 G in plot (b) and 100 G in plot (c), respectively. Details see text.}
	\label{fig:osc}
\end{figure}

Fig.\ref{fig_osca}, Fig.\ref{fig_oscb},Fig.\ref{fig_oscc} shows the measured $f(t)$ under an applied TF of 0 G, 50 G and 100 G, respectively. 
The open black dots with error bars are the data and the red lines are the fits to the data, whereas the blue dashed lines are the exponential component acquired after the fitting and deducting the oscillating component.
The fitting results are listed in table \ref{tab:fitresult}.

\begin{table}[h]
\centering
\caption{Result of Fitting the data measured at the three Magnetic Field Conditions}.
\begin{tabular}{c|c|c|c}
\Xhline{1.2pt}
TF (G) & 0 & 50 & 100 \\
\Xhline{1.2pt}
$\tau_\mu$ (ns) & $2189 \pm 15$ & $2203 \pm 7$ & $2200 \pm 5$ \\
\hline
$B_\mu$ (G) & -- & $52.2 \pm 0.1$ & $104.0 \pm 0.1$ \\
\hline
$A_0$ & -- & $0.245 \pm 0.004$ & $0.251 \pm 0.004$ \\
\hline
$N_0$ & $1649 \pm 12$ & $6800 \pm 91$ & $4985 \pm 24$ \\
\hline
$C$ & $-0.15 \pm 1.71$ & $-3.36 \pm 3.25$ & $-0.10 \pm 0.18$ \\ 
\hline
$\chi^2/NDF$ & 0.95 & 1.08 & 0.93 \\
\Xhline{1.2pt}
\end{tabular}
\label{tab:fitresult}
\end{table}

Fig.\ref{fig_osca} is fitted with an exponential function as expressed by (\ref{eq:eq0}):

\begin{equation}
    f(t)=N_0 \cdot e^{- \frac{t}{\tau_\mu}} + C
    \label{eq:eq0}
\end{equation}

where, $N_0$ represents the starting intensity of muon decay. C represents a common background correction factor.
$\tau_\mu$ represents the lifetime of muon and is fitted as 2.2 $\mu$s, which validates that the LGAD performs well at detecting the decay positrons of muons considering the coincidence to the expected theoretical value.

In Fig.\ref{fig_oscb} and Fig.\ref{fig_oscc}, the fitting is performed using the asymmetry functions \cite{muSRBook-ISIS} as expressed by (\ref{eq:eq1}).

\begin{equation}
    f(t)=N_0 \cdot e^{- \frac{t}{\tau_\mu}} \cdot (1+A_0\cdot P_i(t))+C
    \label{eq:eq1}
\end{equation}

where, $A_0$ is the asymmetry factor depending on the covered solid angle of the detector and $P_i(t)$ is the norm projection of the muon polarization along the direction of the detector.
$P_i(t)$ is calculated from the formula expressed by (\ref{eq:eq2})

\begin{equation}
    P_i(t)=\mathrm{cos^2}\theta + \mathrm{sin^2}\theta\cdot \mathrm{cos}(\gamma_\mu B_\mu t)
    \label{eq:eq2}
\end{equation}

where, $\gamma_\mu$ is the gyro-magnetic ratio of muon, with constant value of $2\pi\times135.5 MHz/T$, $B_\mu$ is the magnetic field seen by the muons and $\theta$ represents the angle of the magnetic field with respect to the initial polarization.
In this test, a TF was set in a perpendicular direction to the muon beam (as shown in Fig.\ref{fig_device}).
Thus, (\ref{eq:eq2}) would be simplified to (\ref{eq:eq3})

\begin{equation}
    P_i(t)=\mathrm{cos}(\gamma_\mu B_\mu t)
    \label{eq:eq3}
\end{equation}

Considering a common background correction factor C and a timing bias correction factor b, the muon decay time detected by the LGAD can be expressed by:

\begin{equation}
    f(t)=N_0 \cdot e^{- \frac{t}{\tau_\mu}} \cdot (1+A_0\cdot \mathrm{cos}(\gamma_\mu B_\mu (t-b)))+C
    \label{eq:eq4}
\end{equation}

\begin{figure}[h]
    \centering
    \subfigure[]{    
		\label{fig_asyma}     
		\includegraphics[width=0.4\textwidth]{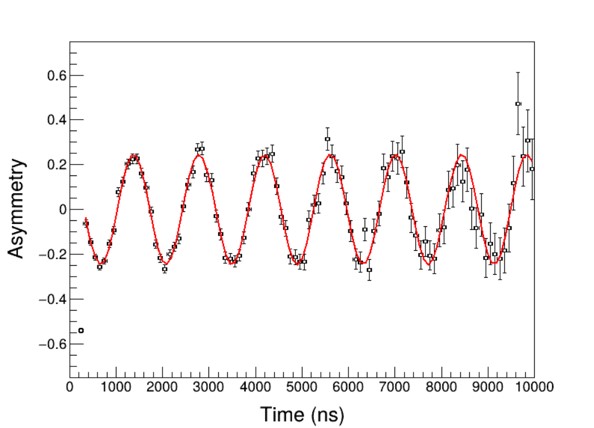}}
      \qquad
    \subfigure[]{    
		\label{fig_asymb}     
		\includegraphics[width=0.4\textwidth]{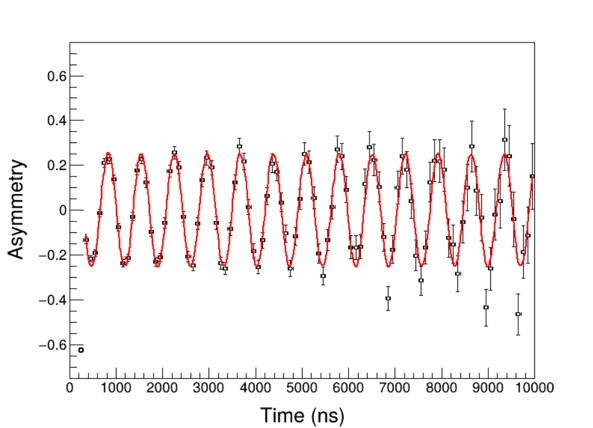}}
	\caption{Asymmetry signals in 50 G (left panel) and 100 G (right panel), after correction for the exponential decay.}
	\label{fig:asym}
\end{figure}

With the formula in (\ref{eq:eq4}), the data in Fig.\ref{fig_oscb} and Fig.\ref{fig_oscc} are fitted.
Actually, the \msr signals shown in Fig.\ref{fig:osc} and expressed in (\ref{eq:eq4}) have two dominant components, the decay background and the asymmetry signal. 
The decaying background component is contributed by the nature of muon decays, which obey an exponential distribution with a mean lifetime of 2.2 $\mu$s.
The asymmetry signal, which is a cosine function in time, corresponds to the Larmor precession of the muon spin.
This precessional component has been separated out and shown in Fig.\ref{fig:asym} after the fitting analysis.
The amplitude $A_0$ (the asymmetry factor) is fitted to be around 0.25 under the field configurations of 50 G and 100 G.
This value is coincident to our expectation since no degrader were used during the test and the detector is almost on the plane of precessing spins.
The angular frequency, depending on the magnetic field $B_\mu$, are also different under both field configurations.
With the fitting, $B_\mu$ are found to be 52 G and 104 G.
Considering the coincidence of the offset between the applied and measured fields in both cases implies that the magnetic field generated by the CHRONUS magnet had a general offset of around 4\% during our test.


\subsection{Discussions}

During the test, the event rate received by LGAD was stable at around 2.3 events per second (or around 0.06 events per muon pulse), which is very low.
This is because of two reasons.
One reason is that the covering angle of the the LGAD used for the test was small.
The other reason is that the intensity of the muon source (level of $10^5$ muons per second) was not large enough to test the high rate capability of the LGAD.
Another test of a scintillator detector, which is the current detector design for application to the new \msr spectrometer on MELODY, was conducted in parallel with this experiment on the LGAD device.
In the scintillator test, the event rate was of the level of 1 event per detector per pulse, which is around 17 times the event rate seen in the LGAD test.
Given the fact that the pulse width of the LGAD is around 2 ns and the segmentation size of millimeters, the utilized intensity of the muon beam is far less than the upper limit that can be detected by the LGAD.
Thus, considering the features of fast in time, the capability of LGAD when detecting the \msr signal is much larger than the scintillator.

In summary, comparing to the "scintillator with SiPM" applied in conventional \msr spectrometers, the advantages of the LGAD are as follows.
Firstly, the LGAD is much faster, which would break the limitation of dead time of the SiPM and improve the allowed counting rate of each single detector element.
Secondly, the LGAD is much smaller in size, which makes it possible to construct a detector matrix complex with a high channel density and achieve a higher overall counting rate in the spectrometer.
Thirdly, the active volume of the LGAD is much thinner than the scintillator, which would naturally reduce the prevalence of multi-counting across neighboring detectors.
Additionally, the LGAD has a high signal-to-noise ratio since it is not sensitive to gammas when compared to the scintillator. 
Therefore, the gamma annihilation background would be much decreased.
Based on the advantages described above, the LGAD is expected to be a good candidate for the next generation of detectors used in \msr spectrometers.


\section{Conclusion and Prospects}
\label{sec:conclusion}

A test of LGAD technology has been carried out using the CHRONUS instrument at the ISIS muon source, RAL, UK.
The details of the experiment setup as well as the subsequent data analysis have been discussed in this paper, from which several conclusions can be presented.

In the first, considering the coincidence between the measured muon life time and the expected theoretical value, the LGAD is able to detect the very fast decay positrons from muon decays, which is vital in deducing the evolution of muon spin polarization in a \msr experiment. 
Secondly, given the fact that the \msr signals are clearly observed for a test sample in different transverse magnetic fields, the data of the LGAD have provided explicit evidence of the Larmor precession of muon spins. From this signal, the magnetic fields in the sample have been extracted and found to correspond to the expected applied magnetic field.
Thirdly, based on the fits to the asymmetry data, the asymmetry term as function of time can be explicitly separated from the muon decay lifetime signal, which is essential in analyzing the magnetic properties of materials in future \msr experiments.
Finally, considering the detector features of fast response and small pixel size, the speed of the LGAD in the detection of \msr signals is much faster than the scintillator system. This suggests that the LGAD has the potential to enable full utilization of more intense pulsed muon sources and increase the efficiency of \msr experiments, which is especially relevant to a low repetition rate source such as MELODY.  

Consequently, with the test of the LGAD technology and the subsequent analysis discussed in this paper, the LGAD is proven to be a promising option for the design of next generation \msr spectrometers. The LGAD is especially expected to greatly improve the performance of future \msr spectrometers at the MELODY facility, allowing improved utilization of high intensity muon beams.

\end{document}